\documentclass[%
 aip,
 apl,%
 amsmath,amssymb,
 reprint,%
]{revtex4-1}

\usepackage{graphics}
\usepackage{subfigure}
\usepackage{dcolumn}
\usepackage{bm}

\begin{document}

\preprint{AIP/123-QED}

\title{Perfect Spin-filtering and Giant Magnetoresistance with Fe-terminated Graphene Nanoribbon}

\author{Chao Cao}
 \email{ccao@hznu.edu.cn}
 \affiliation{
 Department of Physics, Hangzhou Normal University, Hangzhou, Zhejiang Province 310036, P. R. China}
\author{Yan Wang}
 \affiliation{Department of Physics and Quantum Theory Project, University of Florida, Gainesville, FL 32611, U. S. A.}%

\author{Hai-Ping Cheng}%
 \affiliation{Department of Physics and Quantum Theory Project, University of Florida, Gainesville, FL 32611, U. S. A.}%
 \email{cheng@qtp.ufl.edu}
 
\author{Jianzhong Jiang}
 \affiliation{International Center for New Structured Materials (ICNSM), Laboratory of New Structured Materials (LNSM), Department of Materials Science and Engineering, Zhejiang University, Hangzhou 310013, People’s Republic of China}

\date{\today}

\begin{abstract}
Spin-dependent electronic transport properties of Fe-terminated zig-zag graphene nanoribbons (zGNR) have been studied using first-principles transport simulations. The spin configuration of proposed zGNR junction can be controlled with external magnetic field, and the tunneling junction show MR$>$1000 at small bias and is a perfect spin-filter by applying uniform external magnetic filed at small bias. 
\end{abstract}

\maketitle


Graphene, a single layer of graphite honeycomb lattice, has attracted great attention since its discovery since 2004\cite{Novoselov22102004}. The material exhibits peculiar electronic structure and large mean free path, and has been therefore considered a potential replacement for silicon in the future electronics\cite{nmat_6_183}. Son {\it et al.} studied the electronic structure of graphene nanoribbons (GNR), and proposed that the zig-zag GNR (zGNR) serve as electrically controllable spin-valve\cite{ISI:000242018300043}. This result quickly draws attention and the study of GNRs becomes the focus of the graphene research. Various methods to prepare GNRs or to enhance their qualities have been proposed\cite{ISI:000280141200031,ISI:000268138600016,ISI:000265182500039,ISI:000257984700010,ISI:000253530600039}; the edge states and stabilities have been explored\cite{ISI:000265479300049,ISI:000259195800042,ISI:000254408000021,ISI:000251678700070,ISI:000245512400041,ISI:000242786500020,ISI:000238696600133};  the doping and edge-termination properties have been studied\cite{ISI:000255262900020,ISI:000254408000021,ISI:000253530600039,ISI:000246413200041,ISI:000245512400041,ISI:000245329600023,ISI:000284400700004}; and possible applications in electronic devices have been discussed\cite{acr4_43_111,ISI:000263815800087,ISI:000263166400042,ISI:000257984700012,ISI:000251908100089,ISI:000248595800094,ISI:000247186800006,ISI:000246909900074,ISI:000242100200085,ISI:000284400700004}. In this letter, we show a possible implementation of GNR based spintronics device, which shows large magnetoresistance (MR) as well as perfect spin-filtering effect.

\begin{figure}[htp]
  \centering
  \scalebox{0.8}{\includegraphics{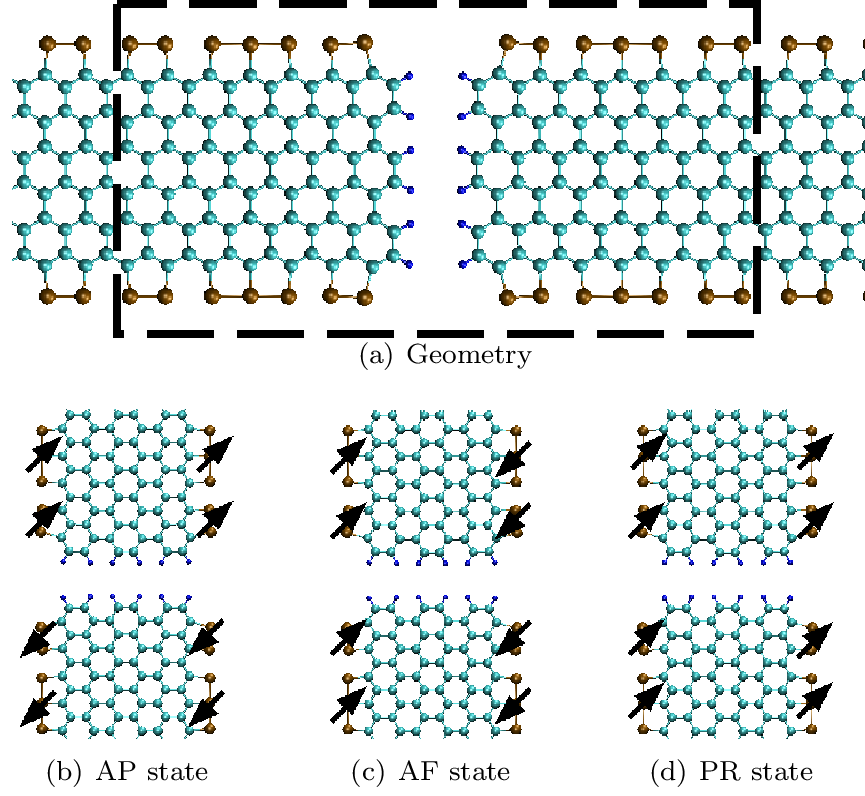}}
  \caption{(color online)Geometry (a) and spin configurations (b-d) of the model system. The dashed line enclosed region is the scattering region in simulations. The arrows on the atoms indicates the direction of local magnetic moment.\label{fig_structure}}
\end{figure}

The proposed tunneling junction is modelled with two semi-infinite Fe-terminated zGNRs with hydrogen termination at the ends (Fig. \ref{fig_structure}(a)). Previous studies have shown that the transition metal terminated zGNRs (TM-zGNR) exhibits both antiferromagnetic (AFM) and ferromagnetic (FM) long-range ordering like the prestine zGNRs and the hydrogen-terminated zGNRs (H-zGNR)\cite{ISI:000284400700004}. Therefore, at least three distinct spin states can be realized with the proposed junction: 1. edges within each lead are ferromagnetically coupled, while edges across the leads are antiferromagnetically coupled (Fig. \ref{fig_structure}(b)); 2. edges within each lead are antiferromagnetically coupled, while edges across the leads are ferromagnetically coupled (Fig. \ref{fig_structure}(c)); 3. edges within each leads as well as across the leads are ferromagnetically coupled (Fig. \ref{fig_structure}(d)). In the following text, these states will be referred to as the AP state, the AF state, and the PR state, respectively. Experimentally, the AF state is the ground state without any external $\mathrm{B}$-field, while an uniform external $\mathrm{B}$-field will turn the system into the PR state. The AP state could be induced if external $\mathrm{B}$-fields in opposite direction are applied to the two leads. 

In order to model the transport properties of the proposed tunneling junction at different states, we employed non-equilibrium Green's function (NEGF) method combined with density functional theory (DFT). The structure of the modelled system was fully relaxed using the plane-wave basis DFT code PWSCF\cite{PWSCF_code}, with the ultrasoft pseudopotential method. The energy cutoffs for plane-wave and density were chosen to be 40 Ry and 320 Ry, respectively. The transport properties was then obtained with the optimized geometry using the SIESTA code\cite{SIESTA_1,*SIESTA_2}, with the norm-conserving pseudopotential method and atomic orbitals. The real-space mesh grid size was chosen to represent 200 Ry equivalent plane-wave energy cutoff, and single-$\zeta$ basis with optimized parameters were employed to avoid the singularity problems in calculating the lead Green's function. In all calculations, Perdew, Berke, and Ernzerhoff parameterization (PBE) to general gradient approximation (GGA)\cite{PBE} was used to calculate the electron exchange-correlation. 

\begin{figure}
 \centering
 \rotatebox{270}{\scalebox{0.44}{\includegraphics{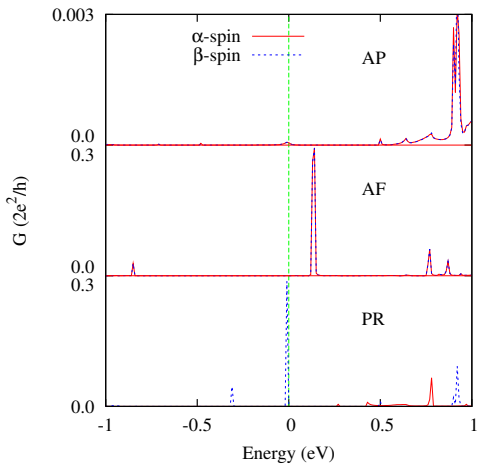}}}
 \caption{(color online) Transmission coefficient G(E) of the tunneling junction at AP (upper panel), AF (middle panel), and PR (lower panel) states at zero-bias. Please be noted that the conductance scale of the upper panel is two magnitudes larger than the other two. The Fermi level is shifted to 0 (the green dashed line).
 \label{fig_te}
 }
\end{figure}

We first analyze the transport properties of this GNR junction. We show the transmission coefficient $G(E)$ of the GNR junction at different states in Fig. \ref{fig_te}. Due to the symmetry of the junction and the spacial spin configuration, the $\alpha$-spin and $\beta$-spin channels are degenerate for both AP and AF states, as shown in the figure. For the tunneling junction at the PR or AF states, since both leads are exactly the same, the electron band energies of both leads in either spin channel perfectly match each other, and therefore the transmission coefficient $G(E)$ of these states are two magnitudes larger than $G(E)$ of the AP state. As the state of the tunneling junction could be controlled with external magnetic field, it is therefore expected that the tunneling junction will show significant magnetoresistivity behavior. For the tunneling junction at the PR state, it can further be noticed that the degeneracy over spin channels is lifted, and a large conductance peak shows up at the Fermi level $E_F$ for the $\beta$-spin channel, while the $\alpha$-spin channel does not open until around 0.4 eV. Thus, it is expected that the junction could serve as a perfect spin-filter at PR state.

\begin{figure}[htp]
 \centering
 \scalebox{0.89}{\includegraphics{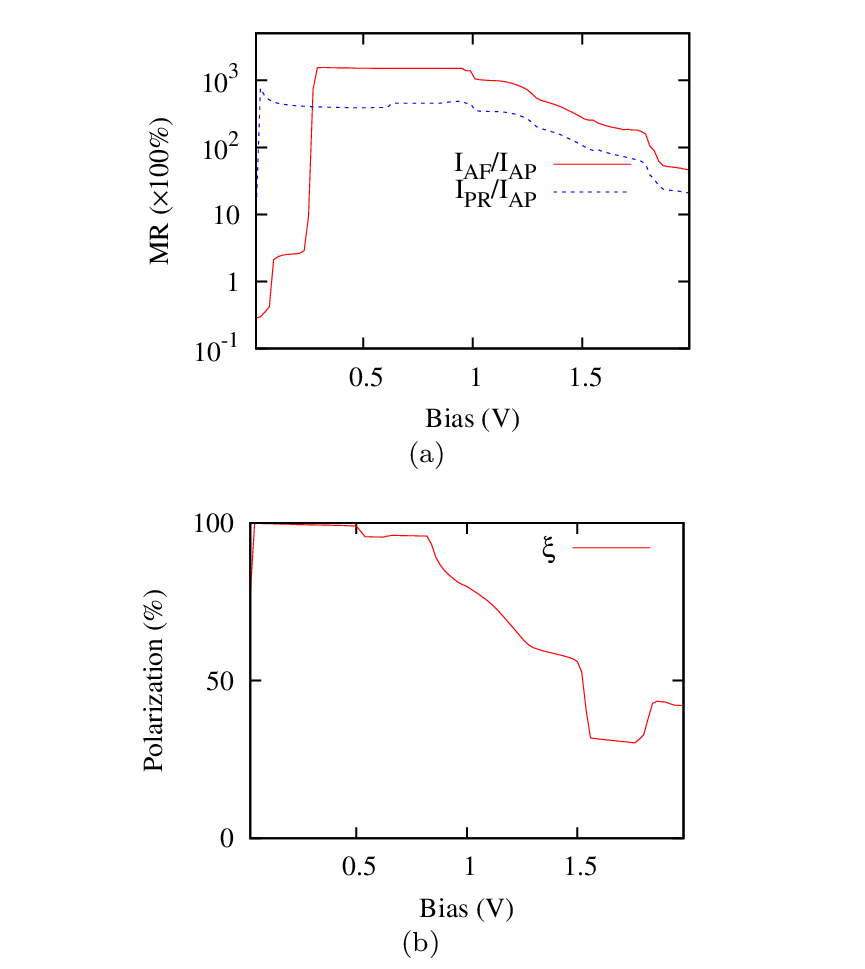}}
 \caption{a) Magnetoresistance of PR/AP and AF/AP states; and b) current-polarization at PR state. The $y$-axis in panel a) is in logarithm scale.
 \label{fig_mr_ip}
 }
\end{figure}

To further illustrate the possible MR and spin-filtering behavior, we employed the following equation
$$I=\int_{-\infty}^{\infty} \frac{2e}{h}G(E)\left(f_L(E)-f_R(E)\right)dE$$
to calculate the current at small bias. The obtained current is then used to calculate the MR between PR/AP and AF/AP states (FIG. \ref{fig_mr_ip}(a)). As expected, the junction shows MR order of 10$^{3}$ between either PR/AP states or AF/AP states for a wide range of bias voltage. In the AF/AP case, the small conductance hump around $E_F$ at the AP state causes the MR initially smaller than 1. However, as the conductance peak of AF state is close to $E_F$ and its scale is two magnitudes larger, the current at AP state quickly become dominant and the MR reaches over 1000 from $\approx 0.3$ to 1 V. For the junction at the PR state, the obtained spin-current is also used to calculate the current polarization $\xi$ defined with
$$\xi=\frac{I_{\beta}-I_{\alpha}}{I_{\alpha}+I_{\beta}}$$
where $\alpha$ and $\beta$ denotes majority and minority spin, respectively. As shown in FIG. \ref{fig_mr_ip}(b), the spin polarization of the current is almost 100\% for small bias ($V<0.5$ V).

\begin{figure*}[htp]
 \centering
 \scalebox{0.9}{\includegraphics{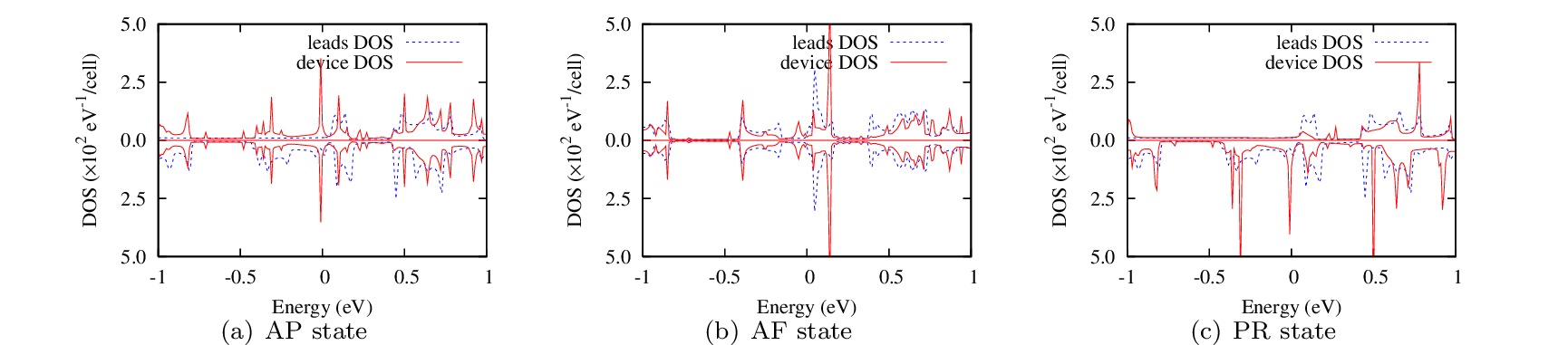}}
 \caption{Density of states of {\it left} lead and scattering region at a) AP, b) AF and c) PR states. The upper panels are for $\alpha$-spin; while the lower panels are for $\beta$-spin. 
 \label{fig_dos}
 }
\end{figure*}

In order to understand the physics behind the GMR and spin-filtering behavior of the Fe-terminated GNR, we have performed density of states (DOS) calculations. In either AP or AF states, both spin channels of the scattering region are degenerate as shown in FIG. \ref{fig_dos}(a) and \ref{fig_dos}(b) due to the spacial symmetry, in consistency with the transmission coefficient data. For the tunneling junction at AP state, although the scattering region DOS show large peaks around $E_F$ for both spin channels, the left/right lead DOS are almost negligible for $\alpha$-/$\beta$-spin respectively (the figure shows only the left lead DOS, the right lead DOS is the same except that the spin channels needs to be swapped). Furthermore, the spin transfer for this state is also forbidden due to the mismatch between orbital symmetries at both sides, similar to the case in the pure zGNR device\cite{acr4_43_111}. Therefore, neither spin can easily tunnel through the junction, and thus the transmission coefficient at AP state is orders of magnitudes lower than other two. The scattering region DOS at AF state show large peaks around 0.1 eV, where small humps are also present in the leads DOS. This feature implies resonant tunneling through the junction, which leads to the large transmission coefficient peak at AF state around 0.1 eV (middle panel of FIG. \ref{fig_te}). Finally, for the tunneling junction at the PR state, both leads and the scattering region have the same orbital symmetry as well as aligned energy levels\cite{acr4_43_111}, thus a perfect transmission is expected. However, as both leads and scattering region show negligible DOS for the $\alpha$-spin channel\ref{fig_dos}(c), therefore the $\alpha$-spin conductance is greatly suppressed. In contrast, for the $\beta$-spin channel, the scattering region DOS show a large peak at $E_F$, where a small DOS hump is also present for the leads DOS. Thus, resonance tunneling takes place at small bias for the $\beta$-spin. Hence, the junction show significant spin-filtering effect at small bias.

In conclusion, we have performed first-principles transport simulations on Fe-terminated GNR junctions, whose spin configurations may be controlled by applying external magnetic field. The junction show MR over 1000 at small bias for a large bias range, and is a perfect spin-filter at PR state at small bias. The behavior of the junction is controlled by the DOS of both leads and scattering region as well as the associated orbital/band symmetries.

This work is supported by National Science Foundation of China NSFC No. 10904127 (C. Cao) and U.S./DOE/BES under Grant No. DE-FG02-02ER45995 (H.-P. Cheng and Y. Wang). All calculations were performed at the High Performance Computing Center of Hangzhou Normal University.


\begin{thebibliography}{31}%
\makeatletter
\providecommand \@ifxundefined [1]{%
 \@ifx{#1\undefined}
}%
\providecommand \@ifnum [1]{%
 \ifnum #1\expandafter \@firstoftwo
 \else \expandafter \@secondoftwo
 \fi
}%
\providecommand \@ifx [1]{%
 \ifx #1\expandafter \@firstoftwo
 \else \expandafter \@secondoftwo
 \fi
}%
\providecommand \natexlab [1]{#1}%
\providecommand \enquote  [1]{``#1''}%
\providecommand \bibnamefont  [1]{#1}%
\providecommand \bibfnamefont [1]{#1}%
\providecommand \citenamefont [1]{#1}%
\providecommand \href@noop [0]{\@secondoftwo}%
\providecommand \href [0]{\begingroup \@sanitize@url \@href}%
\providecommand \@href[1]{\@@startlink{#1}\@@href}%
\providecommand \@@href[1]{\endgroup#1\@@endlink}%
\providecommand \@sanitize@url [0]{\catcode `\\12\catcode `\$12\catcode
  `\&12\catcode `\#12\catcode `\^12\catcode `\_12\catcode `\%12\relax}%
\providecommand \@@startlink[1]{}%
\providecommand \@@endlink[0]{}%
\providecommand \url  [0]{\begingroup\@sanitize@url \@url }%
\providecommand \@url [1]{\endgroup\@href {#1}{\urlprefix }}%
\providecommand \urlprefix  [0]{URL }%
\providecommand \Eprint [0]{\href }%
\providecommand \doibase [0]{http://dx.doi.org/}%
\providecommand \selectlanguage [0]{\@gobble}%
\providecommand \bibinfo  [0]{\@secondoftwo}%
\providecommand \bibfield  [0]{\@secondoftwo}%
\providecommand \translation [1]{[#1]}%
\providecommand \BibitemOpen [0]{}%
\providecommand \bibitemStop [0]{}%
\providecommand \bibitemNoStop [0]{.\EOS\space}%
\providecommand \EOS [0]{\spacefactor3000\relax}%
\providecommand \BibitemShut  [1]{\csname bibitem#1\endcsname}%
\let\auto@bib@innerbib\@empty
\bibitem [{\citenamefont {Novoselov}\ \emph {et~al.}(2004)\citenamefont
  {Novoselov}, \citenamefont {Geim}, \citenamefont {Morozov}, \citenamefont
  {Jiang}, \citenamefont {Zhang}, \citenamefont {Dubonos}, \citenamefont
  {Grigorieva},\ and\ \citenamefont {Firsov}}]{Novoselov22102004}%
  \BibitemOpen
  \bibfield  {author} {\bibinfo {author} {\bibfnamefont {K.~S.}\ \bibnamefont
  {Novoselov}}, \bibinfo {author} {\bibfnamefont {A.~K.}\ \bibnamefont {Geim}},
  \bibinfo {author} {\bibfnamefont {S.~V.}\ \bibnamefont {Morozov}}, \bibinfo
  {author} {\bibfnamefont {D.}~\bibnamefont {Jiang}}, \bibinfo {author}
  {\bibfnamefont {Y.}~\bibnamefont {Zhang}}, \bibinfo {author} {\bibfnamefont
  {S.~V.}\ \bibnamefont {Dubonos}}, \bibinfo {author} {\bibfnamefont {I.~V.}\
  \bibnamefont {Grigorieva}}, \ and\ \bibinfo {author} {\bibfnamefont {A.~A.}\
  \bibnamefont {Firsov}},\ }\href@noop {} {\bibfield  {journal} {\bibinfo
  {journal} {Science}\ }\textbf {\bibinfo {volume} {306}},\ \bibinfo {pages}
  {666} (\bibinfo {year} {2004})}\BibitemShut {NoStop}%
\bibitem [{\citenamefont {Geim}\ and\ \citenamefont
  {Novoselov}(2007)}]{nmat_6_183}%
  \BibitemOpen
  \bibfield  {author} {\bibinfo {author} {\bibfnamefont {A.~K.}\ \bibnamefont
  {Geim}}\ and\ \bibinfo {author} {\bibfnamefont {K.~S.}\ \bibnamefont
  {Novoselov}},\ }\href@noop {} {\bibfield  {journal} {\bibinfo  {journal}
  {Nat. Mater.}\ }\textbf {\bibinfo {volume} {6}},\ \bibinfo {pages} {183}
  (\bibinfo {year} {2007})}\BibitemShut {NoStop}%
\bibitem [{\citenamefont {Son}, \citenamefont {Cohen},\ and\ \citenamefont
  {Louie}(2006)}]{ISI:000242018300043}%
  \BibitemOpen
  \bibfield  {author} {\bibinfo {author} {\bibfnamefont {Y.-W.}\ \bibnamefont
  {Son}}, \bibinfo {author} {\bibfnamefont {M.~L.}\ \bibnamefont {Cohen}}, \
  and\ \bibinfo {author} {\bibfnamefont {S.~G.}\ \bibnamefont {Louie}},\
  }\href@noop {} {\bibfield  {journal} {\bibinfo  {journal} {Nature}\ }\textbf
  {\bibinfo {volume} {444}},\ \bibinfo {pages} {347} (\bibinfo {year}
  {2006})}\BibitemShut {NoStop}%
\bibitem [{\citenamefont {Cai}\ \emph {et~al.}(2010)\citenamefont {Cai},
  \citenamefont {Ruffieux}, \citenamefont {Jaafar}, \citenamefont {Bieri},
  \citenamefont {Braun}, \citenamefont {Blankenburg}, \citenamefont {Muoth},
  \citenamefont {Seitsonen}, \citenamefont {Saleh}, \citenamefont {Feng},
  \citenamefont {Muellen},\ and\ \citenamefont {Fasel}}]{ISI:000280141200031}%
  \BibitemOpen
  \bibfield  {author} {\bibinfo {author} {\bibfnamefont {J.}~\bibnamefont
  {Cai}}, \bibinfo {author} {\bibfnamefont {P.}~\bibnamefont {Ruffieux}},
  \bibinfo {author} {\bibfnamefont {R.}~\bibnamefont {Jaafar}}, \bibinfo
  {author} {\bibfnamefont {M.}~\bibnamefont {Bieri}}, \bibinfo {author}
  {\bibfnamefont {T.}~\bibnamefont {Braun}}, \bibinfo {author} {\bibfnamefont
  {S.}~\bibnamefont {Blankenburg}}, \bibinfo {author} {\bibfnamefont
  {M.}~\bibnamefont {Muoth}}, \bibinfo {author} {\bibfnamefont {A.~P.}\
  \bibnamefont {Seitsonen}}, \bibinfo {author} {\bibfnamefont {M.}~\bibnamefont
  {Saleh}}, \bibinfo {author} {\bibfnamefont {X.}~\bibnamefont {Feng}},
  \bibinfo {author} {\bibfnamefont {K.}~\bibnamefont {Muellen}}, \ and\
  \bibinfo {author} {\bibfnamefont {R.}~\bibnamefont {Fasel}},\ }\href@noop {}
  {\bibfield  {journal} {\bibinfo  {journal} {Nature}\ }\textbf {\bibinfo
  {volume} {466}},\ \bibinfo {pages} {470} (\bibinfo {year}
  {2010})}\BibitemShut {NoStop}%
\bibitem [{\citenamefont {Campos}\ \emph {et~al.}(2009)\citenamefont {Campos},
  \citenamefont {Manfrinato}, \citenamefont {Sanchez-Yamagishi}, \citenamefont
  {Kong},\ and\ \citenamefont {Jarillo-Herrero}}]{ISI:000268138600016}%
  \BibitemOpen
  \bibfield  {author} {\bibinfo {author} {\bibfnamefont {L.~C.}\ \bibnamefont
  {Campos}}, \bibinfo {author} {\bibfnamefont {V.~R.}\ \bibnamefont
  {Manfrinato}}, \bibinfo {author} {\bibfnamefont {J.~D.}\ \bibnamefont
  {Sanchez-Yamagishi}}, \bibinfo {author} {\bibfnamefont {J.}~\bibnamefont
  {Kong}}, \ and\ \bibinfo {author} {\bibfnamefont {P.}~\bibnamefont
  {Jarillo-Herrero}},\ }\href@noop {} {\bibfield  {journal} {\bibinfo
  {journal} {Nano Lett.}\ }\textbf {\bibinfo {volume} {9}},\ \bibinfo {pages}
  {2600} (\bibinfo {year} {2009})}\BibitemShut {NoStop}%
\bibitem [{\citenamefont {Kosynkin}\ \emph {et~al.}(2009)\citenamefont
  {Kosynkin}, \citenamefont {Higginbotham}, \citenamefont {Sinitskii},
  \citenamefont {Lomeda}, \citenamefont {Dimiev}, \citenamefont {Price},\ and\
  \citenamefont {Tour}}]{ISI:000265182500039}%
  \BibitemOpen
  \bibfield  {author} {\bibinfo {author} {\bibfnamefont {D.~V.}\ \bibnamefont
  {Kosynkin}}, \bibinfo {author} {\bibfnamefont {A.~L.}\ \bibnamefont
  {Higginbotham}}, \bibinfo {author} {\bibfnamefont {A.}~\bibnamefont
  {Sinitskii}}, \bibinfo {author} {\bibfnamefont {J.~R.}\ \bibnamefont
  {Lomeda}}, \bibinfo {author} {\bibfnamefont {A.}~\bibnamefont {Dimiev}},
  \bibinfo {author} {\bibfnamefont {B.~K.}\ \bibnamefont {Price}}, \ and\
  \bibinfo {author} {\bibfnamefont {J.~M.}\ \bibnamefont {Tour}},\ }\href@noop
  {} {\bibfield  {journal} {\bibinfo  {journal} {Nature}\ }\textbf {\bibinfo
  {volume} {458}},\ \bibinfo {pages} {872} (\bibinfo {year}
  {2009})}\BibitemShut {NoStop}%
\bibitem [{\citenamefont {Tapaszto}\ \emph {et~al.}(2008)\citenamefont
  {Tapaszto}, \citenamefont {Dobrik}, \citenamefont {Lambin},\ and\
  \citenamefont {Biro}}]{ISI:000257984700010}%
  \BibitemOpen
  \bibfield  {author} {\bibinfo {author} {\bibfnamefont {L.}~\bibnamefont
  {Tapaszto}}, \bibinfo {author} {\bibfnamefont {G.}~\bibnamefont {Dobrik}},
  \bibinfo {author} {\bibfnamefont {P.}~\bibnamefont {Lambin}}, \ and\ \bibinfo
  {author} {\bibfnamefont {L.~P.}\ \bibnamefont {Biro}},\ }\href@noop {}
  {\bibfield  {journal} {\bibinfo  {journal} {Nat. Nanotech.}\ }\textbf
  {\bibinfo {volume} {3}},\ \bibinfo {pages} {397} (\bibinfo {year}
  {2008})}\BibitemShut {NoStop}%
\bibitem [{\citenamefont {Li}\ \emph {et~al.}(2008)\citenamefont {Li},
  \citenamefont {Wang}, \citenamefont {Zhang}, \citenamefont {Lee},\ and\
  \citenamefont {Dai}}]{ISI:000253530600039}%
  \BibitemOpen
  \bibfield  {author} {\bibinfo {author} {\bibfnamefont {X.}~\bibnamefont
  {Li}}, \bibinfo {author} {\bibfnamefont {X.}~\bibnamefont {Wang}}, \bibinfo
  {author} {\bibfnamefont {L.}~\bibnamefont {Zhang}}, \bibinfo {author}
  {\bibfnamefont {S.}~\bibnamefont {Lee}}, \ and\ \bibinfo {author}
  {\bibfnamefont {H.}~\bibnamefont {Dai}},\ }\href@noop {} {\bibfield
  {journal} {\bibinfo  {journal} {Science}\ }\textbf {\bibinfo {volume}
  {319}},\ \bibinfo {pages} {1229} (\bibinfo {year} {2008})}\BibitemShut
  {NoStop}%
\bibitem [{\citenamefont {Huang}\ \emph {et~al.}(2009)\citenamefont {Huang},
  \citenamefont {Liu}, \citenamefont {Su}, \citenamefont {Wu}, \citenamefont
  {Duan}, \citenamefont {Gu},\ and\ \citenamefont {Liu}}]{ISI:000265479300049}%
  \BibitemOpen
  \bibfield  {author} {\bibinfo {author} {\bibfnamefont {B.}~\bibnamefont
  {Huang}}, \bibinfo {author} {\bibfnamefont {M.}~\bibnamefont {Liu}}, \bibinfo
  {author} {\bibfnamefont {N.}~\bibnamefont {Su}}, \bibinfo {author}
  {\bibfnamefont {J.}~\bibnamefont {Wu}}, \bibinfo {author} {\bibfnamefont
  {W.}~\bibnamefont {Duan}}, \bibinfo {author} {\bibfnamefont {B.-l.}\
  \bibnamefont {Gu}}, \ and\ \bibinfo {author} {\bibfnamefont {F.}~\bibnamefont
  {Liu}},\ }\href@noop {} {\bibfield  {journal} {\bibinfo  {journal} {Phys.
  Rev. Lett.}\ }\textbf {\bibinfo {volume} {102}},\ \bibinfo {pages} {166404}
  (\bibinfo {year} {2009})}\BibitemShut {NoStop}%
\bibitem [{\citenamefont {Wassmann}\ \emph {et~al.}(2008)\citenamefont
  {Wassmann}, \citenamefont {Seitsonen}, \citenamefont {Saitta}, \citenamefont
  {Lazzeri},\ and\ \citenamefont {Mauri}}]{ISI:000259195800042}%
  \BibitemOpen
  \bibfield  {author} {\bibinfo {author} {\bibfnamefont {T.}~\bibnamefont
  {Wassmann}}, \bibinfo {author} {\bibfnamefont {A.~P.}\ \bibnamefont
  {Seitsonen}}, \bibinfo {author} {\bibfnamefont {A.~M.}\ \bibnamefont
  {Saitta}}, \bibinfo {author} {\bibfnamefont {M.}~\bibnamefont {Lazzeri}}, \
  and\ \bibinfo {author} {\bibfnamefont {F.}~\bibnamefont {Mauri}},\
  }\href@noop {} {\bibfield  {journal} {\bibinfo  {journal} {Phys. Rev. Lett.}\
  }\textbf {\bibinfo {volume} {101}},\ \bibinfo {pages} {096402} (\bibinfo
  {year} {2008})}\BibitemShut {NoStop}%
\bibitem [{\citenamefont {Kudin}(2008)}]{ISI:000254408000021}%
  \BibitemOpen
  \bibfield  {author} {\bibinfo {author} {\bibfnamefont {K.~N.}\ \bibnamefont
  {Kudin}},\ }\href@noop {} {\bibfield  {journal} {\bibinfo  {journal} {ACS
  Nano}\ }\textbf {\bibinfo {volume} {2}},\ \bibinfo {pages} {516} (\bibinfo
  {year} {2008})}\BibitemShut {NoStop}%
\bibitem [{\citenamefont {Kan}\ \emph {et~al.}(2007)\citenamefont {Kan},
  \citenamefont {Li}, \citenamefont {Yang},\ and\ \citenamefont
  {Hou}}]{ISI:000251678700070}%
  \BibitemOpen
  \bibfield  {author} {\bibinfo {author} {\bibfnamefont {E.-J.}\ \bibnamefont
  {Kan}}, \bibinfo {author} {\bibfnamefont {Z.}~\bibnamefont {Li}}, \bibinfo
  {author} {\bibfnamefont {J.}~\bibnamefont {Yang}}, \ and\ \bibinfo {author}
  {\bibfnamefont {J.~G.}\ \bibnamefont {Hou}},\ }\href@noop {} {\bibfield
  {journal} {\bibinfo  {journal} {Appl. Phys. Lett.}\ }\textbf {\bibinfo
  {volume} {91}},\ \bibinfo {pages} {243116} (\bibinfo {year}
  {2007})}\BibitemShut {NoStop}%
\bibitem [{\citenamefont {Jiang}, \citenamefont {Sumpter},\ and\ \citenamefont
  {Dai}(2007)}]{ISI:000245512400041}%
  \BibitemOpen
  \bibfield  {author} {\bibinfo {author} {\bibfnamefont {D.-e.}\ \bibnamefont
  {Jiang}}, \bibinfo {author} {\bibfnamefont {B.~G.}\ \bibnamefont {Sumpter}},
  \ and\ \bibinfo {author} {\bibfnamefont {S.}~\bibnamefont {Dai}},\
  }\href@noop {} {\bibfield  {journal} {\bibinfo  {journal} {J. Chem. Phys.}\
  }\textbf {\bibinfo {volume} {126}},\ \bibinfo {pages} {134701} (\bibinfo
  {year} {2007})}\BibitemShut {NoStop}%
\bibitem [{\citenamefont {Barone}, \citenamefont {Hod},\ and\ \citenamefont
  {Scuseria}(2006)}]{ISI:000242786500020}%
  \BibitemOpen
  \bibfield  {author} {\bibinfo {author} {\bibfnamefont {V.}~\bibnamefont
  {Barone}}, \bibinfo {author} {\bibfnamefont {O.}~\bibnamefont {Hod}}, \ and\
  \bibinfo {author} {\bibfnamefont {G.~E.}\ \bibnamefont {Scuseria}},\
  }\href@noop {} {\bibfield  {journal} {\bibinfo  {journal} {Nano Lett.}\
  }\textbf {\bibinfo {volume} {6}},\ \bibinfo {pages} {2748} (\bibinfo {year}
  {2006})}\BibitemShut {NoStop}%
\bibitem [{\citenamefont {Brey}\ and\ \citenamefont
  {Fertig}(2006)}]{ISI:000238696600133}%
  \BibitemOpen
  \bibfield  {author} {\bibinfo {author} {\bibfnamefont {L.}~\bibnamefont
  {Brey}}\ and\ \bibinfo {author} {\bibfnamefont {H.}~\bibnamefont {Fertig}},\
  }\href@noop {} {\bibfield  {journal} {\bibinfo  {journal} {Phys. Rev. B}\
  }\textbf {\bibinfo {volume} {73}},\ \bibinfo {pages} {235411} (\bibinfo
  {year} {2006})}\BibitemShut {NoStop}%
\bibitem [{\citenamefont {Kim},\ and\ \citenamefont {Kim}(2010)}]{acr4_43_111}%
  \BibitemOpen
  \bibfield  {author} {\bibinfo {author} {\bibfnamefont {W.~Y.}\ \bibnamefont
  {Kim}},\ and\ \bibinfo {author} {\bibfnamefont {K.~S.} \bibnamefont{Kim}},\ }\href
  {\doibase 10.1021/ar900156u} {\bibfield {journal} {\bibinfo {journal} {Acc. Chem.
  Res.}\ }\textbf {\bibinfo {volume} {43}},\ \bibinfo{pages} {111} (\bibinfo {year} 
  {2010})}\BibitemShut {Nostop}%
\bibitem [{\citenamefont {Yu}\ \emph {et~al.}(2008)\citenamefont {Yu},
  \citenamefont {Zheng}, \citenamefont {Wen},\ and\ \citenamefont
  {Jiang}}]{ISI:000255262900020}%
  \BibitemOpen
  \bibfield  {author} {\bibinfo {author} {\bibfnamefont {S.~S.}\ \bibnamefont
  {Yu}}, \bibinfo {author} {\bibfnamefont {W.~T.}\ \bibnamefont {Zheng}},
  \bibinfo {author} {\bibfnamefont {Q.~B.}\ \bibnamefont {Wen}}, \ and\
  \bibinfo {author} {\bibfnamefont {Q.}~\bibnamefont {Jiang}},\ }\href@noop {}
  {\bibfield  {journal} {\bibinfo  {journal} {Carbon}\ }\textbf {\bibinfo
  {volume} {46}},\ \bibinfo {pages} {537} (\bibinfo {year} {2008})}\BibitemShut
  {NoStop}%
\bibitem [{\citenamefont {Martins}\ \emph {et~al.}(2007)\citenamefont
  {Martins}, \citenamefont {Miwa}, \citenamefont {da~Silva},\ and\
  \citenamefont {Fazzio}}]{ISI:000246413200041}%
  \BibitemOpen
  \bibfield  {author} {\bibinfo {author} {\bibfnamefont {T.~B.}\ \bibnamefont
  {Martins}}, \bibinfo {author} {\bibfnamefont {R.~H.}\ \bibnamefont {Miwa}},
  \bibinfo {author} {\bibfnamefont {A.~J.~R.}\ \bibnamefont {da~Silva}}, \ and\
  \bibinfo {author} {\bibfnamefont {A.}~\bibnamefont {Fazzio}},\ }\href@noop {}
  {\bibfield  {journal} {\bibinfo  {journal} {Phys. Rev. Lett.}\ }\textbf
  {\bibinfo {volume} {98}},\ \bibinfo {pages} {196803} (\bibinfo {year}
  {2007})}\BibitemShut {NoStop}%
\bibitem [{\citenamefont {Wang}\ \emph
  {et~al.}(2007{\natexlab{a}})\citenamefont {Wang}, \citenamefont {Li},
  \citenamefont {Zheng}, \citenamefont {Ren}, \citenamefont {Su}, \citenamefont
  {Shi},\ and\ \citenamefont {Chen}}]{ISI:000245329600023}%
  \BibitemOpen
  \bibfield  {author} {\bibinfo {author} {\bibfnamefont {Z.~F.}\ \bibnamefont
  {Wang}}, \bibinfo {author} {\bibfnamefont {Q.}~\bibnamefont {Li}}, \bibinfo
  {author} {\bibfnamefont {H.}~\bibnamefont {Zheng}}, \bibinfo {author}
  {\bibfnamefont {H.}~\bibnamefont {Ren}}, \bibinfo {author} {\bibfnamefont
  {H.}~\bibnamefont {Su}}, \bibinfo {author} {\bibfnamefont {Q.~W.}\
  \bibnamefont {Shi}}, \ and\ \bibinfo {author} {\bibfnamefont
  {J.}~\bibnamefont {Chen}},\ }\href@noop {} {\bibfield  {journal} {\bibinfo
  {journal} {Phys. Rev. B}\ }\textbf {\bibinfo {volume} {75}},\ \bibinfo
  {pages} {113406} (\bibinfo {year} {2007}{\natexlab{a}})}\BibitemShut
  {NoStop}%
\bibitem [{\citenamefont {Wang}, \citenamefont {Cao},\ and\ \citenamefont
  {Cheng}(2010)}]{ISI:000284400700004}%
  \BibitemOpen
  \bibfield  {author} {\bibinfo {author} {\bibfnamefont {Y.}~\bibnamefont
  {Wang}}, \bibinfo {author} {\bibfnamefont {C.}~\bibnamefont {Cao}}, \ and\
  \bibinfo {author} {\bibfnamefont {H.-P.}\ \bibnamefont {Cheng}},\ }\href@noop
  {} {\bibfield  {journal} {\bibinfo  {journal} {Phys. Rev. B}\ }\textbf
  {\bibinfo {volume} {82}},\ \bibinfo {pages} {205429} (\bibinfo {year}
  {2010})}\BibitemShut {NoStop}%
\bibitem [{\citenamefont {Mucciolo}, \citenamefont {Castro~Neto},\ and\
  \citenamefont {Lewenkopf}(2009)}]{ISI:000263815800087}%
  \BibitemOpen
  \bibfield  {author} {\bibinfo {author} {\bibfnamefont {E.~R.}\ \bibnamefont
  {Mucciolo}}, \bibinfo {author} {\bibfnamefont {A.~H.}\ \bibnamefont
  {Castro~Neto}}, \ and\ \bibinfo {author} {\bibfnamefont {C.~H.}\ \bibnamefont
  {Lewenkopf}},\ }\href@noop {} {\bibfield  {journal} {\bibinfo  {journal}
  {Phys. Rev. B}\ }\textbf {\bibinfo {volume} {79}},\ \bibinfo {pages} {075407}
  (\bibinfo {year} {2009})}\BibitemShut {NoStop}%
\bibitem [{\citenamefont {Stampfer}\ \emph {et~al.}(2009)\citenamefont
  {Stampfer}, \citenamefont {Gutttinger}, \citenamefont {Hellmueller},
  \citenamefont {Molitor}, \citenamefont {Ensslin},\ and\ \citenamefont
  {Ihn}}]{ISI:000263166400042}%
  \BibitemOpen
  \bibfield  {author} {\bibinfo {author} {\bibfnamefont {C.}~\bibnamefont
  {Stampfer}}, \bibinfo {author} {\bibfnamefont {J.}~\bibnamefont
  {Gutttinger}}, \bibinfo {author} {\bibfnamefont {S.}~\bibnamefont
  {Hellmueller}}, \bibinfo {author} {\bibfnamefont {F.}~\bibnamefont
  {Molitor}}, \bibinfo {author} {\bibfnamefont {K.}~\bibnamefont {Ensslin}}, \
  and\ \bibinfo {author} {\bibfnamefont {T.}~\bibnamefont {Ihn}},\ }\href@noop
  {} {\bibfield  {journal} {\bibinfo  {journal} {Phys. Rev. Lett.}\ }\textbf
  {\bibinfo {volume} {102}},\ \bibinfo {pages} {056403} (\bibinfo {year}
  {2009})}\BibitemShut {NoStop}%
\bibitem [{\citenamefont {Kim}\ and\ \citenamefont
  {Kim}(2008)}]{ISI:000257984700012}%
  \BibitemOpen
  \bibfield  {author} {\bibinfo {author} {\bibfnamefont {W.~Y.}\ \bibnamefont
  {Kim}}\ and\ \bibinfo {author} {\bibfnamefont {K.~S.}\ \bibnamefont {Kim}},\
  }\href@noop {} {\bibfield  {journal} {\bibinfo  {journal} {Nat. Nanotech.}\
  }\textbf {\bibinfo {volume} {3}},\ \bibinfo {pages} {408} (\bibinfo {year}
  {2008})}\BibitemShut {NoStop}%
\bibitem [{\citenamefont {Huang}\ \emph {et~al.}(2007)\citenamefont {Huang},
  \citenamefont {Yan}, \citenamefont {Zhou}, \citenamefont {Wu}, \citenamefont
  {Gu}, \citenamefont {Duan},\ and\ \citenamefont {Liu}}]{ISI:000251908100089}%
  \BibitemOpen
  \bibfield  {author} {\bibinfo {author} {\bibfnamefont {B.}~\bibnamefont
  {Huang}}, \bibinfo {author} {\bibfnamefont {Q.}~\bibnamefont {Yan}}, \bibinfo
  {author} {\bibfnamefont {G.}~\bibnamefont {Zhou}}, \bibinfo {author}
  {\bibfnamefont {J.}~\bibnamefont {Wu}}, \bibinfo {author} {\bibfnamefont
  {B.-L.}\ \bibnamefont {Gu}}, \bibinfo {author} {\bibfnamefont
  {W.}~\bibnamefont {Duan}}, \ and\ \bibinfo {author} {\bibfnamefont
  {F.}~\bibnamefont {Liu}},\ }\href@noop {} {\bibfield  {journal} {\bibinfo
  {journal} {Appl. Phys. Lett.}\ }\textbf {\bibinfo {volume} {91}},\ \bibinfo
  {pages} {253122} (\bibinfo {year} {2007})}\BibitemShut {NoStop}%
\bibitem [{\citenamefont {Wang}\ \emph
  {et~al.}(2007{\natexlab{b}})\citenamefont {Wang}, \citenamefont {Shi},
  \citenamefont {Li}, \citenamefont {Wang}, \citenamefont {Hou}, \citenamefont
  {Zheng}, \citenamefont {Yao},\ and\ \citenamefont
  {Chen}}]{ISI:000248595800094}%
  \BibitemOpen
  \bibfield  {author} {\bibinfo {author} {\bibfnamefont {Z.~F.}\ \bibnamefont
  {Wang}}, \bibinfo {author} {\bibfnamefont {Q.~W.}\ \bibnamefont {Shi}},
  \bibinfo {author} {\bibfnamefont {Q.}~\bibnamefont {Li}}, \bibinfo {author}
  {\bibfnamefont {X.}~\bibnamefont {Wang}}, \bibinfo {author} {\bibfnamefont
  {J.~G.}\ \bibnamefont {Hou}}, \bibinfo {author} {\bibfnamefont
  {H.}~\bibnamefont {Zheng}}, \bibinfo {author} {\bibfnamefont
  {Y.}~\bibnamefont {Yao}}, \ and\ \bibinfo {author} {\bibfnamefont
  {J.}~\bibnamefont {Chen}},\ }\href@noop {} {\bibfield  {journal} {\bibinfo
  {journal} {Appl. Phys. Lett.}\ }\textbf {\bibinfo {volume} {91}},\ \bibinfo
  {pages} {053109} (\bibinfo {year} {2007}{\natexlab{b}})}\BibitemShut
  {NoStop}%
\bibitem [{\citenamefont {Yan}\ \emph {et~al.}(2007)\citenamefont {Yan},
  \citenamefont {Huang}, \citenamefont {Yu}, \citenamefont {Zheng},
  \citenamefont {Zang}, \citenamefont {Wu}, \citenamefont {Gu}, \citenamefont
  {Liu},\ and\ \citenamefont {Duan}}]{ISI:000247186800006}%
  \BibitemOpen
  \bibfield  {author} {\bibinfo {author} {\bibfnamefont {Q.}~\bibnamefont
  {Yan}}, \bibinfo {author} {\bibfnamefont {B.}~\bibnamefont {Huang}}, \bibinfo
  {author} {\bibfnamefont {J.}~\bibnamefont {Yu}}, \bibinfo {author}
  {\bibfnamefont {F.}~\bibnamefont {Zheng}}, \bibinfo {author} {\bibfnamefont
  {J.}~\bibnamefont {Zang}}, \bibinfo {author} {\bibfnamefont {J.}~\bibnamefont
  {Wu}}, \bibinfo {author} {\bibfnamefont {B.-L.}\ \bibnamefont {Gu}}, \bibinfo
  {author} {\bibfnamefont {F.}~\bibnamefont {Liu}}, \ and\ \bibinfo {author}
  {\bibfnamefont {W.}~\bibnamefont {Duan}},\ }\href@noop {} {\bibfield
  {journal} {\bibinfo  {journal} {Nano Lett.}\ }\textbf {\bibinfo {volume}
  {7}},\ \bibinfo {pages} {1469} (\bibinfo {year} {2007})}\BibitemShut
  {NoStop}%
\bibitem [{\citenamefont {Xu}, \citenamefont {Zheng},\ and\ \citenamefont
  {Chen}(2007)}]{ISI:000246909900074}%
  \BibitemOpen
  \bibfield  {author} {\bibinfo {author} {\bibfnamefont {Z.}~\bibnamefont
  {Xu}}, \bibinfo {author} {\bibfnamefont {Q.-S.}\ \bibnamefont {Zheng}}, \
  and\ \bibinfo {author} {\bibfnamefont {G.}~\bibnamefont {Chen}},\ }\href@noop
  {} {\bibfield  {journal} {\bibinfo  {journal} {Appl. Phys. Lett.}\ }\textbf
  {\bibinfo {volume} {90}},\ \bibinfo {pages} {223115} (\bibinfo {year}
  {2007})}\BibitemShut {NoStop}%
\bibitem [{\citenamefont {Ouyang}\ \emph {et~al.}(2006)\citenamefont {Ouyang},
  \citenamefont {Yoon}, \citenamefont {Fodor},\ and\ \citenamefont
  {Guo}}]{ISI:000242100200085}%
  \BibitemOpen
  \bibfield  {author} {\bibinfo {author} {\bibfnamefont {Y.}~\bibnamefont
  {Ouyang}}, \bibinfo {author} {\bibfnamefont {Y.}~\bibnamefont {Yoon}},
  \bibinfo {author} {\bibfnamefont {J.~K.}\ \bibnamefont {Fodor}}, \ and\
  \bibinfo {author} {\bibfnamefont {J.}~\bibnamefont {Guo}},\ }\href@noop {}
  {\bibfield  {journal} {\bibinfo  {journal} {Appl. Phys. Lett.}\ }\textbf
  {\bibinfo {volume} {89}},\ \bibinfo {pages} {203107} (\bibinfo {year}
  {2006})}\BibitemShut {NoStop}%
\bibitem [{\citenamefont {Giannozzi}\ \emph {et~al.}(2009)\citenamefont
  {Giannozzi}, \citenamefont {Baroni}, \citenamefont {Bonini},\ and\
  \citenamefont {{\it et al.}}}]{PWSCF_code}%
  \BibitemOpen
  \bibfield  {author} {\bibinfo {author} {\bibfnamefont {P.}~\bibnamefont
  {Giannozzi}}, \bibinfo {author} {\bibfnamefont {S.}~\bibnamefont {Baroni}},
  \bibinfo {author} {\bibfnamefont {N.}~\bibnamefont {Bonini}}, \ and\ \bibinfo
  {author} {\bibnamefont {{\it et al.}}},\ }\href
  {http://www.quantum-espresso.org} {\bibfield  {journal} {\bibinfo  {journal}
  {J. Phys. Cond. Mat.}\ }\textbf {\bibinfo {volume} {21}},\ \bibinfo {pages}
  {395502 (19pp)} (\bibinfo {year} {2009})}\BibitemShut {NoStop}%
\bibitem [{\citenamefont {Soler}\ \emph {et~al.}(2002)\citenamefont {Soler},
  \citenamefont {Artacho}, \citenamefont {Gale}, \citenamefont {García},
  \citenamefont {Junquera}, \citenamefont {Ordejón},\ and\ \citenamefont
  {Sánchez-Portal}}]{SIESTA_1}%
  \BibitemOpen
  \bibfield  {author} {\bibinfo {author} {\bibfnamefont {J.~M.}\ \bibnamefont
  {Soler}}, \bibinfo {author} {\bibfnamefont {E.}~\bibnamefont {Artacho}},
  \bibinfo {author} {\bibfnamefont {J.~D.}\ \bibnamefont {Gale}}, \bibinfo
  {author} {\bibfnamefont {A.}~\bibnamefont {García}}, \bibinfo {author}
  {\bibfnamefont {J.}~\bibnamefont {Junquera}}, \bibinfo {author}
  {\bibfnamefont {P.}~\bibnamefont {Ordejón}}, \ and\ \bibinfo {author}
  {\bibfnamefont {D.}~\bibnamefont {Sánchez-Portal}},\ }\href
  {http://stacks.iop.org/0953-8984/14/i=11/a=302} {\bibfield  {journal}
  {\bibinfo  {journal} {J. Phys. Cond. Mat.}\ }\textbf {\bibinfo {volume}
  {14}},\ \bibinfo {pages} {2745} (\bibinfo {year} {2002})}\BibitemShut
  {NoStop}%
\bibitem [{\citenamefont {Brandbyge}\ \emph {et~al.}(2002)\citenamefont
  {Brandbyge}, \citenamefont {Mozos}, \citenamefont {Ordej\'on}, \citenamefont
  {Taylor},\ and\ \citenamefont {Stokbro}}]{SIESTA_2}%
  \BibitemOpen
  \bibfield  {author} {\bibinfo {author} {\bibfnamefont {M.}~\bibnamefont
  {Brandbyge}}, \bibinfo {author} {\bibfnamefont {J.-L.}\ \bibnamefont
  {Mozos}}, \bibinfo {author} {\bibfnamefont {P.}~\bibnamefont {Ordej\'on}},
  \bibinfo {author} {\bibfnamefont {J.}~\bibnamefont {Taylor}}, \ and\ \bibinfo
  {author} {\bibfnamefont {K.}~\bibnamefont {Stokbro}},\ }\href {\doibase
  10.1103/PhysRevB.65.165401} {\bibfield  {journal} {\bibinfo  {journal} {Phys.
  Rev. B}\ }\textbf {\bibinfo {volume} {65}},\ \bibinfo {pages} {165401}
  (\bibinfo {year} {2002})}\BibitemShut {NoStop}%
\bibitem [{\citenamefont {Perdew}, \citenamefont {Burke},\ and\ \citenamefont
  {Ernzerhof}(1996)}]{PBE}%
  \BibitemOpen
  \bibfield  {author} {\bibinfo {author} {\bibfnamefont {J.~P.}\ \bibnamefont
  {Perdew}}, \bibinfo {author} {\bibfnamefont {K.}~\bibnamefont {Burke}}, \
  and\ \bibinfo {author} {\bibfnamefont {M.}~\bibnamefont {Ernzerhof}},\ }\href
  {\doibase 10.1103/PhysRevLett.77.3865} {\bibfield  {journal} {\bibinfo
  {journal} {Phys. Rev. Lett.}\ }\textbf {\bibinfo {volume} {77}},\ \bibinfo
  {pages} {3865} (\bibinfo {year} {1996})}\BibitemShut {NoStop}%
\end{thebibliography}

%

\end{document}